\documentclass[epj]{svjour}
\usepackage{latexsym}
\usepackage{graphicx}
\usepackage{xcolor}
\begin{document}
\newcommand{\eg}{{\it e.g.}}
\newcommand{\etal}{{\it et. al.}}
\newcommand{\ie}{{\it i.e.}}
\newcommand{\be}{\begin{equation}}
\newcommand{\dd}{\displaystyle}
\newcommand{\ee}{\end{equation}}
\newcommand{\bea}{\begin{eqnarray}}
\newcommand{\eea}{\end{eqnarray}}
\newcommand{\bef}{\begin{figure}}
\newcommand{\eef}{\end{figure}}
\newcommand{\bce}{\begin{center}}
\newcommand{\ece}{\end{center}}
\def\lsim{\mathrel{\rlap{\lower4pt\hbox{\hskip1pt$\sim$}}
    \raise1pt\hbox{$<$}}}         
\def\gsim{\mathrel{\rlap{\lower4pt\hbox{\hskip1pt$\sim$}}
    \raise1pt\hbox{$>$}}}         
\title{Hybrid star structure with the Field Correlator Method }
\author{G. F. Burgio\inst{1} \and D. Zappal\`a\inst{1}
}                     
\offprints{}          
\institute{INFN Sezione di Catania, Via Santa Sofia 64, 95123 Catania, Italy 
\phantom { \and INFN Sezione di Catania, Via Santa Sofia 64, 95123 Catania, Italy}
}
\date{Received: date / Revised version: date}
%
\abstract{We explore the relevance of the color-flavor locking phase in the equation of state (EoS) built 
with the Field Correlator Method (FCM) for the description of the quark matter core of hybrid stars. 
For the hadronic phase, we use the microscopic Brueckner-Hartree-Fock (BHF) many-body theory, 
and its relativistic counterpart, i.e. the Dirac-Brueckner (DBHF). 
We find that the main features of the phase transition are directly related to the values of the quark-antiquark
potential $V_1$, the gluon condensate $G_2$ and the color-flavor superconducting gap $\Delta$.
We confirm that the mapping between the FCM and the CSS (constant speed of sound) parameterization holds true even in the
case of paired quark matter. The inclusion of hyperons in the hadronic phase and its effect on the mass-radius relation of hybrid stars
is also investigated. 
} 
\maketitle

\section{Introduction}
\label{sec:1}

The possible appearance of quark matter (QM) in the interior of massive neutron stars (NS)
is currently one of the main theoretical issues in the physics of compact stars \cite{2000Glen}. 
The existence of two NS of about two solar masses has been confirmed by 
recent observations  \cite{2010Natur:Demorest,Antoniadis:2013en} . Based on a microscopic nucleonic equation of state (EoS),
one expects that in such heavy NS the central particle density reaches values larger than $1/{fm}^3$,
where in fact quark degrees of freedom are expected to appear at a macroscopic level.
Unfortunately, while the microscopic theory of the nucleonic EoS has
reached a high degree of sophistication \cite{Baldo:1997ag,1998APR,Zhou:2004br,Baldo:2012nh,Li:2008bp},
the QM EoS is still poorly known at zero temperature and
at the high baryonic density appropriate for NS.  In fact the essential theoretical tool, i.e. lattice formulation
of the quantum
chromodynamics (QCD) is inapplicable at large baryon densities and small temperature due to the so-called Sign Problem
 \cite{deForc}, and this is due to its complicated nonlinear and nonperturbative nature. 
On the other hand, in the large temperature and small density region lattice
QCD simulations have provided controlled results for the EoS as well as for the 
nature of the transition \cite{2006Aoki,2009Baza}.

The mass of a NS can be calculated by solving the Tolman-Oppenheimer-Volkoff
(TOV) equations with the relevant EoS as input.
The hybrid EoS including both hadronic matter and QM is usually obtained
by combining EoSs of hadronic matter and QM within individual theories/models.

Continuing a set of investigations using different quark models \cite{nsquark,2003njl,2004cdm,Baldo:2006bt,2011Chen},
recently we have explored the nature of the phase transition with the Field Correlator Method (FCM) model of quark matter
\cite{Dosch:1987sk,Dosch:1988ha,Simonov:1987rn},  
 which in principle is able to cover the full temperature-chemical potential plane. 
 In our previous papers \cite{Baldo:2008en,Plumari:2013ira},
we tested the FCM model by comparing the results for the neutron star masses with the existing phenomenology,
which puts strong constraints on the parameters of the model, i.e. the quark-antiquark potential $V_1$
and the gluon condensate $G_2$. 

Recently, we found that the FCM model can be expressed in the
language of the "Constant Speed of Sound" (CSS) parameterization \cite{Alford:2013aca,2015FCM},
and we showed how its parameters can be mapped on to the CSS parameter space.
We remind that the CSS scheme is a general parameterization suitable
for expressing experimental constraints in a model-independent way,
and for classifying  different models of quark matter and establishing  
connections among them. 
It is applicable to high-density equations of state for which: (a) there is a sharp interface
between nuclear matter and quark matter, (b) the speed of sound in the high-density matter is pressure-independent
in the range between the first-order transition pressure up to the maximum central pressure of neutron stars. 
Given the nuclear matter EoS $\epsilon_{\rm NM}(p)$, the high-density EoS can be expressed as 
\begin{equation}
\epsilon(p) = \left\{\!
\begin{array}{ll}
\epsilon_{\rm NM}(p) & p<p_{trans} \\
\epsilon_{\rm NM}(p_{trans})+\Delta\epsilon + c_{\rm QM}^{-2} (p-p_{trans}) & p>p_{trans}
\end{array}
\right.\ 
\label{eqn:CSS_EoS}
\end{equation}
where the three parameters:
the pressure $p_{trans}$ at the transition, the discontinuity in energy
density $\Delta\epsilon$ at the transition, and the speed of sound $c_{QM}$ 
characterize completely the high-density phase. 

In this work we elaborate more on that point by extending the FCM model in order to include
the color superconductivity through the color-flavor locking (CFL) mechanism, which mimics an explicit 
dependence of the gluon condensate $G_2$
on the quark chemical potential, as initially studied in ref.\cite{Plumari:2013ira}. 
We find that the value of the hybrid star maximum mass depends strongly on the FCM model parameters,
i.e. $V_1$, $G_2$ and the gap $\Delta$. We also find that the mapping on to the CSS parameterization 
still holds true when including the CFL phase, though the parameter region explored depends on
the value of the gap $\Delta$. We also pay special attention to the analysis of the hadron-quark 
phase transition when hyperons are included in the hadronic phase. In fact, given the strong softening 
of the EoS due to hyperons,  it has been often found that no quark matter phase transition
can take place \cite{2003njl,2011Chen}.

This paper is organized as follows. In the next section we briefly review the EoS of the hadronic sector, 
particularly the BHF and DBHF microscopic  approaches. In Sect. 3 we discuss the quark matter EoS,
and in Sect. 3.1 we illustrate the FCM at finite
density, with the inclusion of the color-flavor locking effect in Sec. 3.2. Sect. 4 contains numerical results,
with some details of the EoS for neutron star matter and the hadron-quark phase transition in Sect. 4.1.
In Sect. 4.2 we discuss the mass-radius-central density relation for hybrid stars, and the FCM mapping
onto the CSS parameterization. 
Effects due to the inclusion of hyperons are explored in Sect. 4.3. Finally we draw our
conclusions in Sect.5.
 
\section{The BHF and DBHF EoS of Nuclear Matter}
\label{sec:2}

Empirical properties of infinite nuclear matter can be calculated using many different theoretical approaches. 
The amount of experimental and observational data obtained in the last few years, and the 
intense theoretical efforts aimed at their interpretation, call for a firm microscopic approach to the 
modeling of the Equation of State (EoS). In a microscopic approach, the only input required
is a realistic free nucleon-nucleon (NN) interaction with parameters fitted to NN scattering phase shifts in 
different partial wave channels, and to properties of the deuteron. In this paper we adopt
the non-relativistic Brueckner-Bethe-Goldstone (BBG) method \cite{Baldo:1999} and its relativistic counterpart,
the Dirac-Brueckner- Hartree-Fock (DBHF) approximation \cite{GrossBoelting:1998jg}. 

The Brueckner--Bethe--Goldstone (BBG) theory is based on a linked cluster 
expansion of the energy per nucleon of nuclear matter (see Ref.~\cite{Baldo:1999},
chapter 1 and references therein).  
The basic ingredient in this many--body approach is the Brueckner reaction 
matrix $G$, which is the solution of the  Bethe--Goldstone equation 

\begin{equation}
G[\rho;\omega] = v  + \sum_{k_a k_b} v {{|k_a k_b\rangle  Q  \langle k_a k_b|}
  \over {\omega - e(k_a) - e(k_b) }} G[\rho;\omega], 
  \label{eq:G}
\end{equation}                                                           
\noindent
where $v$ is the bare NN interaction, $\rho$ is the nucleon 
number density, and $\omega$ the  starting energy.  
The single-particle energy $e(k)$ (assuming $\hbar$=1),
\begin{equation}
e(k) = e(k;\rho) = {{k^2}\over {2m}} + U(k;\rho),
\label{e:en}
\end{equation}
\noindent
and the Pauli operator $Q$ determine the propagation of intermediate 
baryon pairs. The Brueckner--Hartree--Fock (BHF) approximation for the 
single-particle potential $U$  using the  {\it continuous choice} is
\begin{equation}
U(k;\rho) = \sum _{k'\leq k_F} \langle k k'|G[\rho; e(k)+e(k')]|k k'\rangle_a,
\end{equation}
\noindent
where the subscript ``{\it a}'' indicates antisymmetrization of the 
matrix element.  
Due to the occurrence of $U(k)$ in Eq.~(\ref{e:en}), the above equations constitute 
a coupled system that has to be solved in a self-consistent manner
for several  momenta of the particles involved, at the considered densities.
In the BHF approximation the energy per nucleon is
\begin{equation}
{E \over{A}}  =  
          {{3}\over{5}}{{k_F^2}\over {2m}}  + {{1}\over{2\rho}}  
~ \sum_{k,k'\leq k_F} \langle k k'|G[\rho; e(k)+e(k')]|k k'\rangle_a. 
\end{equation}
\noindent
In this scheme, the only input quantity we need is the bare NN interaction
$v$ in the Bethe-Goldstone equation (1).  It has been shown that the 
nuclear EoS can be 
calculated with good accuracy in the Brueckner two hole-line 
approximation with the continuous choice for the single-particle
potential, since the results in this scheme are quite close to the 
calculations which include also the three hole-line
contribution \cite{Song:1998zz,Baldo:2001mv,Day:1981zz}.  The dependence on the NN interaction, also within
other many-body approaches, has been systematically
investigated in Ref.\cite{Baldo:2012nh}.

However, it is commonly known that  
non-relativistic calculations, based on purely two-body interactions, do not
reproduce correctly the saturation point of symmetric nuclear matter, and that
three-body forces (TBF) are needed to correct this deficiency.
For that, TBF are reduced to a density dependent two-body force by
averaging over the generalized coordinates (position, spin and isospin) of the third particle, 
assuming that the probability of having two particles at a given distance is reduced 
according to the two-body correlation function \cite{Baldo:1997ag,Zhou:2004br,Lovato:2011}.

In this work we will illustrate results obtained using a phenomenological approach to the TBF, 
which is based on  the so-called Urbana model, and consists of an attractive term due to two-pion exchange
with excitation of an intermediate $\Delta$ resonance, and a repulsive 
phenomenological central term \cite{Carlson:1983kq,Schiavilla:1985gb}. 
Within the BHF approach, those TBF produce a shift of about $+1$ MeV in energy and 
$-0.1$ fm$^{-3}$  in density. This adjustment is obtained by tuning the two parameters
contained in the TBF,  and is performed to get an optimal saturation point \cite{Baldo:1997ag,Zhou:2004br}.

Besides a purely phenomenological model, microscopic TBF have also been 
derived
and a tentative approach proposed using the same meson-exchange parameters as the 
underlying NN potential. Results have been obtained with the Argonne $v_{18}$, 
the Bonn B, and the Nijmegen 93 potentials \cite{Li:2008bp,Li:2012zzq}. However, 
at present the theoretical status of microscopically derived TBF is still quite rudimentary, 
Alternatively, latest nuclear matter calculations  \cite{Lovato:2012,Logo:2015} 
used  a new class of chiral inspired TBF \cite{Coon:2001,Epel:2002,Navra:2007}, 
showing that the considered TBF models are not able  to reproduce simultaneously the correct saturation 
point and the properties of three- and four-nucleon systems.

Recently, it has been shown that the role of TBF is greatly reduced if the NN 
potential is based on a realistic constituent quark model \cite{BaldoKenji:2014,Kenji:2015} 
which can explain at the same time few-nucleon systems and nuclear matter, including the
observational data on Neutron Stars and the experimental data on heavy-ions collisions. 
Moreover it has been found that at the highest densities the three-hole-line diagrams can give a
contribution larger than the two-hole-line diagrams,  and this can be related to the characteristic nonlocality
of the repulsive core \cite{Fuji:2013} as produced by the quark-exchange processes.

In the past years, the BHF approach has been extended in order to include the hyperon degrees of freedom
\cite{Baldo:1998hd,Baldo:1999rq}, which play an important role in the
study of neutron star matter.  In fact, hyperons are expected to appear in beta-stable matter already at relatively
low densities of about twice nuclear saturation density, thus producing a softening of the EoS with a strong decrease of the maximum mass. 
If this is the case, the existence of heavy NS would question
the presence of hyperons in their interior, thus requiring alternative
scenarios. It is therefore of great importance to carry out
accurate theoretical calculations of hypernuclear matter starting from 
the available  information on both nucleon and hyperon interactions.
There exist several hyperon-nucleon (NY) potentials fitted
to scattering data, i.e. NSC89 \cite{Maessen:1989}, NSC97 \cite{Stoks:1999}, and ESC08 
\cite{ESC08:2011}, while the hyperon-hyperon (YY) potentials have presently to be considered
rather uncertain or unknown, which is basically due to the lack
of appropriate experimental data. An alternative description of the hyperon-nucleon system has been
recently achieved at next-to-leading order in chiral effective field theory \cite{haiden914,haiden915}.

In the Brueckner scheme, we have used the phenomenological NY potentials \cite{Maessen:1989,Stoks:1999,ESC08:2011} as fundamental input,
and found very low maximum masses of hyperon stars, below 1.4 $M_\odot$ 
($M_\odot = 2 \times 10^{33}\rm g$).  A proposed solution of this so-called {\it hyperon puzzle}
focuses on the role played by hyperonic three-body forces, and several attempts have been made
in this direction \cite{Lona2014,Vidana:2010,Logo2013,Yamamoto:2015lwa}.
However, many inconsistencies still remain, and the solution to this problem is still far from 
being understood.

The relativistic framework is the one on which the nuclear EoS should be
ultimately  based. The best relativistic treatment developed so far is the Dirac-Brueckner (DBHF) 
approach  \cite{GrossBoelting:1998jg}. The DBHF method can be developed in analogy
with the non-relativistic case, i.e. the nucleon inside the nuclear medium is viewed
as a dressed particle in consequence of its two-body interaction with the surrounding
nucleons. The two-body correlations are described by introducing the in-medium
relativistic $G$-matrix. The DBHF scheme can be formulated as a self-consistent problem between the single
particle self-energy $\Sigma$ and the $G$-matrix.  It has been shown that the DBHF
treatment is equivalent \cite{GEBrown:1987} to introducing in the nonrelativistic
BHF the three-body force corresponding to the excitation
of a nucleon-antinucleon pair, the so-called Z-diagram
\cite{1995Baldo}, which is repulsive at all densities,
and consequently produces a saturating effect.  Actually, including in BHF only these particular TBF, 
one gets results close to DBHF calculations, see Ref.\cite{Compilation}.
Generally speaking, the DBHF gives in general a better saturation point  than BHF,  
and the corresponding EoS turns out to be stiffer above saturation than the
one calculated from the BHF + TBF method.
In the relativistic context the only NN potentials which have been developed are the ones
of one-boson exchange type. In the calculations shown here the Bonn A potential is
used \cite{GrossBoelting:1998jg}.
Recently,  the properties of neutron-star matter including hyperons have been investigated within the DBHF approach \cite{kata2015}. In the calculation, the effect of negative-energy states of baryons was partly taken into account, as well as both time and space components of the vector self-energies of baryons and the scalar ones. A value of 2.08 $M_\odot$ was obtained for the maximum neutron-star mass, consistent with the recently observed, massive neutron stars \cite{2010Natur:Demorest,Antoniadis:2013en}.

\section{The quark matter EoS}
\label{sec:3}

\subsection{ The Field Correlator Method }
\label{sec:3_1}

The approach we follow to describe the quark matter EoS
was introduced in \cite{Dosch:1987sk,Dosch:1988ha,Simonov:1987rn} ; see ref.\cite{DiGiacomo:2000va} 
for a review.  For our purposes, we are specifically interested in the 
extension of this approach to finite  baryon  density and temperature,
and all we need is the expression of the pressure as a function of the EoS
thermodynamical  parameters, i.e. the baryon chemical potential  $\mu_B$
and the temperature $T$.  This is derived in \cite{Simonov:2007xc,Simonov:2007jb,Nefediev:2009kn},
and below we report its explicit form . The full pressure, $P_{qg}$, is the sum of the gluon, $P_g$,
the quark, $P_q$,  and the vacuum, $P_v$,  contributions
\be
\label{pqgp1}
P_{qg} =P_g+\sum_{j=u,d,s} P^j_{q} + P_{v}
\ee
where the sum is extended to the three light quark flavors.
The gluon pressure is 
\be
\label{pglue}
P_g = \frac{8 T^4 }{3 \pi^2} \int_0^\infty  d\chi \chi^3
\frac{1}{\exp{(\chi + \frac{9 V_1}{8T} )} - 1}
\ee
while the quark pressure for each single flavor with mass $m_q$ and chemical potential 
$\mu_q$, is 
\be\label{pquark}
P_q= \frac{T^4}{\pi^2}  \left [ \phi_\nu (\frac{\mu_q - V_1/2}{T}) +
\phi_\nu (-\frac{\mu_q + V_1/2} {T}) \right ]
\ee
where 
\be
\label{distrib}
\phi_\nu (a) = \int_0^\infty du \frac{u^4}{\sqrt{u^2+\nu^2}} \frac{1}{(\exp{[ \sqrt{u^2 +
\nu^2} - a]} + 1)},
\ee
being $\nu=m_q/T$. Finally $P_v$, which represents the pressure difference between the 
vacua in the  deconfined and confined phases, is given by 
\be
\label{pqgp2}
P_v= - \frac{(11-\frac{2}{3}N_f)}{32} \frac{G_2}{2}
\ee
where the number of light  flavors in our case is $N_f=3$.

Then, once the quark chemical potentials are related to the baryon chemical 
potential $\mu_B$, the full pressure $P_{qg}$ is defined in terms of
the two parameters $V_1$ and $G_2$ appearing in 
Eqs. (\ref{pglue}-\ref{pqgp2}), where $V_1$  indicates the large distance static $q \bar q$ potential
and  $G_2$ is the gluon condensate.
The former is essentially of nonperturbative nature and  
can be expressed in terms of an integral of a fundamental QCD correlator
\cite{Simonov:2007xc,Simonov:2007jb,Nefediev:2009kn}; however
there is no direct  measurement  of its value. 
The latter is known from QCD sum rules \cite{Shifman:1978bx,Shifman:1978by},
$G_2= < \frac{\alpha_s} {\pi} G^a_{\mu\nu} G^{a \,\mu\nu}> =0.012~ \rm{GeV^4}$,
although an uncertainty of about $50 \%$ affects this estimate.

It is also interesting to notice that $G_2$ appears only in the vacuum contribution to 
the pressure, and  $P_v$ in Eq. (\ref{pqgp2}) has the same role of the bag 
constant of the MIT bag model. Moreover, if one turns off the potential $V_1$,
the quark pressure $P_q$ becomes the pressure of free quarks,  and in this case
the FCM model reduces to the simplest version of the bag model. 
Therefore $V_1$ can be regarded as
the main  correction to the free quarks dynamics inside the bag.
 
In addition to the poor knowledge of the phenomenological values of $V_1$
and $G_2$, one has also to deal with the dependence of these parameters 
on the thermodynamical variables $\mu_B$ and $T$.  
In fact, for the $T$ dependence some indications
can be obtained from the analysis of the deconfinement phase transition 
at $T=T_c$ and $\mu_B=0$,
which is supported by lattice calculations. 
For instance, the fact  that the gluon condensate $G_2$ is 
substantially $T$ independent  except at $T_c$, where it is sharply reduced by one half
\cite{D'Elia:1997ne,D'Elia:2002ck}, 
was already accounted for in the vacuum pressure difference of 
the two phases derived in \cite{Simonov:2007xc}  and reported in Eq. (\ref{pqgp2}).

As far as $V_1$ is concerning, 
the following  expression relating $V_1(T_c)$, $G_2$ and $T_c$ 
is derived within the FCM model, in  \cite{Simonov:2007xc,Simonov:2007jb},
\be
\label{tcritical}
T_c=\frac{a_0 G_2^{1/4}}{2} \left (1 +\sqrt{ 1 + \frac {V_1(T_c)} {2a_0 G_2^{1/4}  }}  \right )\; .
\ee
being $a_0 = (3 \pi^2 / 768)^{1/4}$.
Then,  in  \cite{Bombaci:2012rv} it is shown that,   with $G_2 =0.012 \, \rm{GeV}^4$ 
and  with the lattice estimates of the critical  temperature, $T_c=147\pm 5$  MeV or $T_c=154\pm 9$ MeV,
Eq. (\ref{tcritical}) yields $V_1(T_c)  \lsim 0.15$ GeV.
Another analysis, \cite{Plumari:2013ira}, based on a fit to the lattice determination of the interaction measure $\rm (\epsilon-3p)/T^4$  
at several values of the temperature around and above $T_c$,  suggests a larger potential, $V_1(T_c) \sim 0.5 - 0.6$ GeV, which, according to 
Eq. (\ref{tcritical}) computed at these $T_c$, requires a smaller gluon condensate, $G_2 \simeq 0.003  \div 0.004 \, \rm{GeV^4} $.
In addition to these estimates, an analytic expression for $V_1(T)$, which allows to relate 
$V_1(T_c)$ to the value of the potential at zero temperature, $V_1(T=0)$,
was derived  within the FCM in \cite{Bombaci:2012rv} and, in particular, it turns out that 
$V_1(T_c) \sim 0.5 - 0.6$ GeV corresponds to  $V_1(T=0) \simeq 0.8 \div 0.9$ GeV.

At this point, it is essential to recall that  all these results are obtained at zero baryon density
while, in the core of NS, densities as large as many times the nuclear matter saturation density 
and low temperatures  are expected. Unfortunately, no lattice simulation can be
performed in QCD   at  high  $\mu_B$, and therefore no numerical indication on the
density dependence of  the two parameters  of the FCM is available. On the other hand,
in the extension of the FCM  at finite chemical potential discussed in \cite{Simonov:2007jb},
the authors claim that the potential $V_1$ is expected to be independent of $\mu_B$
at least for small values of $\mu_B$.  However this statement cannot  be straightforwardly  extended 
to  the region of  very large density  where the environment is strongly modified: 
the number of antiparticles becomes much  smaller than the number of particles. 
Therefore, even if the interaction 
strength of quark-antiquark  pairs is larger than  that of quark-quark pairs,
the latter plays a more important role because of the dominance of quarks over antiquarks.  
In addition, within this framework  the particle pairing can lead to the appearance  
of new phases that will be discussed  in Sect. \ref{sec:3_2}.

In view of all the above considerations, 
it is evident that we have very few indications on $V_1$ at large baryon 
density and low temperature, and therefore the best strategy 
is to treat it as a  free parameter of the FCM,  with no {\it a priori} assumed 
dependence on $T$ or $\mu_B$,  which effectively measures 
the large distance interaction strength within a finite quark density  environment, 
and  which  has to be constrained  by the observational data on heaviest NS, through our analysis.
Accordingly, we do not try to relate  $V_1(\mu_B)$  to $V_1(\mu_B=0)$ 
and in particular to those values  of $V_1$ at  deconfinement transition, quoted above.
Therefore,  the widest acceptable range of $V_1$ is explored: so, 
for instance, even  the case  $V_1<0$ is examined, although, as we shall see below, 
it does not produce sufficiently  heavy hybrid NS.
Finally, as explained more in detail in Sect. \ref{sec:3_2},  the other parameter of 
the FCM,  $G_2$, is also treated a free parameter in our analysis, 
essentially for the same reasons discussed in the case of $V_1$.

\subsection{The color-flavor locking effect}
\label{sec:3_2}

In  ref.\cite{Plumari:2013ira} the dependence of $G_2$ on $\mu_B$ was explicitly studied
by introducing a particular ansatz for $G_2(\mu_B)$, which was 
based on previous analysis  of the expectation  of the gluon condensate in dense  
nuclear matter \cite{Cohen:1991nk,Drukarev:2001wd,Baldo:2003id} 
and in two-color, $N_c=2$,  quantum chromodynamics \cite{Metlitski:2005db,Zhitnitsky:2007uk}.
In particular, in \cite{Metlitski:2005db,Zhitnitsky:2007uk} it is shown that  the explicit form obtained for $G_2(\mu_B)$ 
(it starts as a decreasing function at small $\mu_B$ and turns into an increasing 
function at larger $\mu_B$,  with a minimum between these two regions) 
is related to  the effect of diquark pairing and the appearance  of a mass gap.
In fact, at a qualitative level this effect presents strong similarities with 
the occurrence of the color-flavor locking (CFL) superconductive mechanism 
in standard $N_c=3$ QCD \cite{Alford:1998mk,Alford:2007xm}, 
which, as far as the total pressure of 
the quark system is concerned,  induces  an additional  pressure term 
parameterized by a gap $\Delta$. 
Clearly, this new pressure term has an essential role in shaping the $\mu_B$
dependence of the gluon condensate $G_2$, which eventually has the same 
qualitative features of the curve derived in \cite{Zhitnitsky:2007uk}.

Therefore, due to the unavoidable arbitrariness associated to the choice of the 
ansatz for $G_2(\mu_B)$ encountered in \cite{Plumari:2013ira}, rather than following  that approach
we prefer to adopt here 
the same point of view taken for the other FCM parameter
$V_1$, i.e. we take $G_2$ as a free parameter independent of $T$ and $\mu_B$,
that sets  the pressure and the energy density of the vacuum.
Then, the FCM described by the two free parameters $V_1$ and $G_2$, 
is suitable to study  the high density region, and the possible additional contribution 
due to   the new CFL phase, associated to quark-quark pairing, has to be taken into account 
by adding  the additional CFL pressure contribution to the full FCM pressure $P_{qg}$.

The presence of color-flavor locked quark matter is expected at very high $\mu_B$,
and it is realized through quark-quark pairing under the constraint  that the densities 
of the three flavors, up, down and strange, are equal 
\cite{Alford:2007xm,Rajagopal:2000ff,Alford:2002kj,Alford:2004pf}.
The global effect of this pairing on the pressure is the presence of the additional term 
\be\label{cflpress}
P_{cfl}= \frac{\Delta^2\mu_B^2}{3\pi^2}\;\;\; 
\ee
only when  the chemical potential is  greater than $\mu_B=3 m_s^2/(4\Delta)$ and 
the gap $\Delta$ is expected to be in the range $10-100$ MeV in the region of interest 
of $\mu_B$ for the NS. 
Finally the total pressure of the quark matter phase is obtained by adding
$P_{cfl}$ to  $P_{qg}$ given in Eq. (\ref{pqgp1}), and
is treated as a function of the baryon chemical potential $\mu_B$
with three free parameters, namely the potential $V_1$ and the gluon condensate $G_2$, coming from the 
FCM model, and the gap $\Delta$, due to the CFL pairing.

\section{Numerical Results}
\label{sec:4}

\begin{figure*}
\centering
\resizebox{0.82\textwidth}{!}
{\includegraphics*{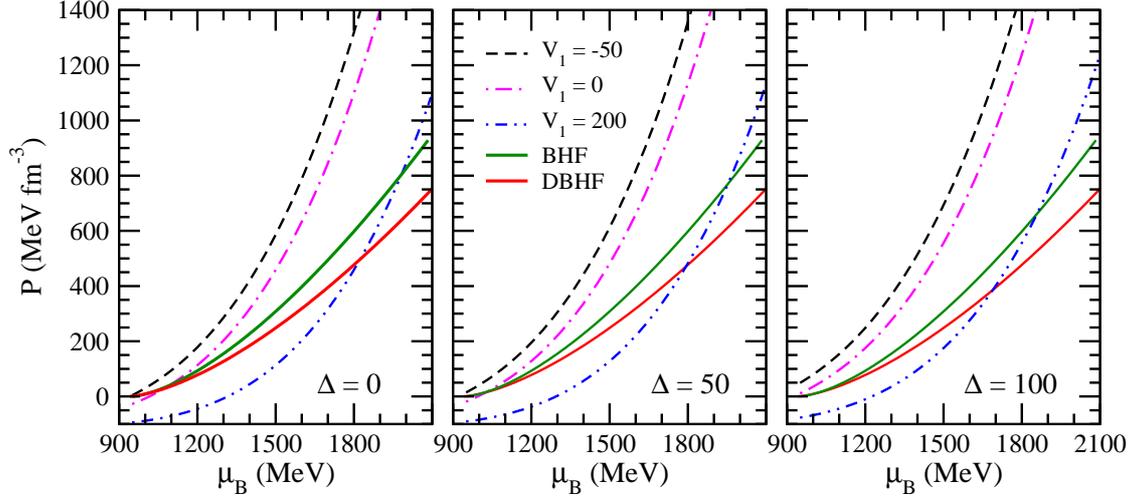}}
\caption{(Color online)The pressure is displayed as a function of the baryon chemical potential 
$\mu_B$ for the FCM quark matter and the purely hadronic matter. All calculations for FCM have been
performed for $G_2 = 0.006$ GeV$^4$, and several values of 
$V_1$ have been chosen. The solid curves represent the BHF (green) and DBHF (red) EoS. Each
panel shows results for different values of the gap $\Delta$, i.e. 0, 50, and 100 MeV. }
\label{fig:1}   
\end{figure*}
\begin{figure*}
\vspace{15mm}
\centering
\resizebox{0.82\textwidth}{!}{%
\includegraphics{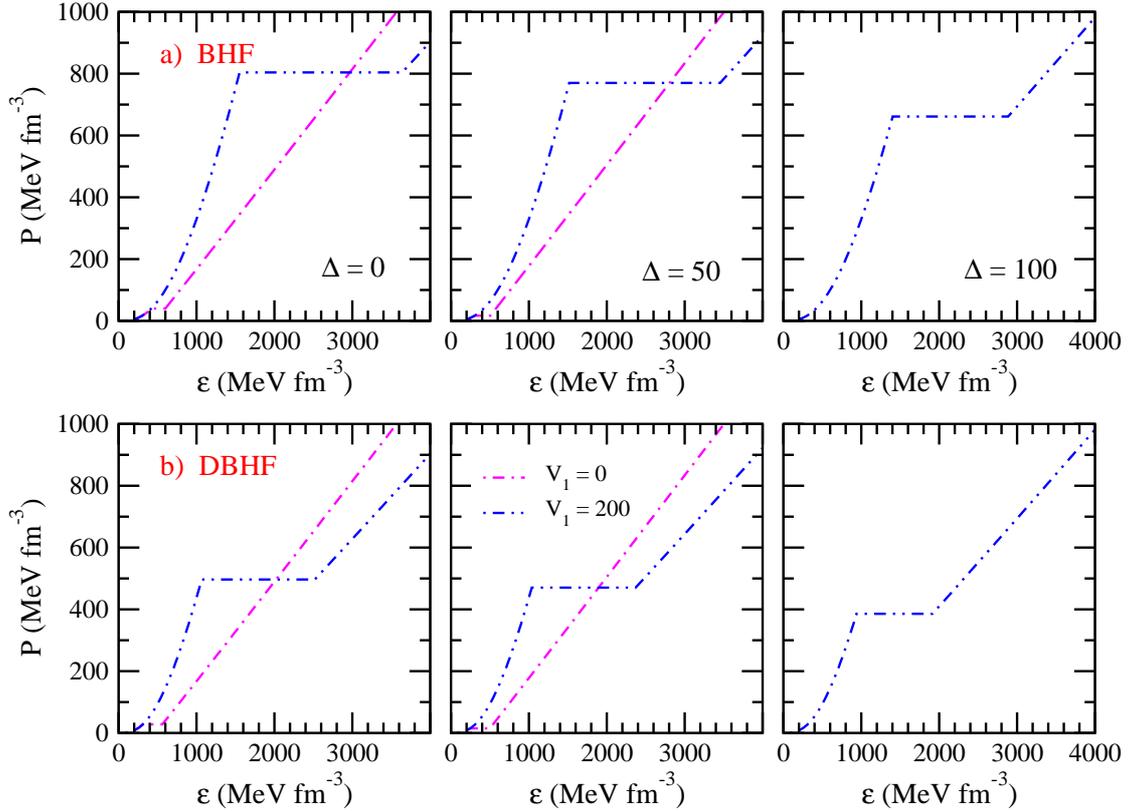}
}
\caption{(Color online)The pressure is displayed vs. the energy density for $V_1 = 0$ (magenta curves) and $V_1 = 200$
(violet) using as hadronic EoS the BHF (upper panels) and the DBHF (lower panels). All calculations 
for FCM have been performed for $G_2 = 0.006$ GeV$^4$. The left, middle and right panels
correspond to the values $\Delta=0, 50, 100$ MeV respectively.}
\label{fig:2}   
\end{figure*}

\subsection{EoS of dense matter in beta equilibrium}
\label{sec:4_1}

In order to study the structure of NS, we have to calculate the composition and the EoS
of cold, neutrino-free, charge-neutral, and beta-stable matter, characterized by two degrees of freedom 
$\mu_B$ and $\mu_e$, the baryon and charge chemical potentials.
The corresponding equations are
\begin{equation}
 \mu_i = b_i \mu_B - q_i \mu_e \:,\quad
 \sum_i \rho_i q_i = 0 \:,
 \label{e:beta}
\end{equation}
$b_i$ and $q_i$ denoting baryon number and charge of
the particle species $i=n,p,e,\mu$ in the hadron phase
and $i=u,d,s,e,\mu$ in the quark phase, respectively.

As far as the hadronic phase is concerning, the Brueckner calculation yields the energy density of
baryon/lepton matter as a function of the different partial densities,
\bea
 \epsilon(\rho_n,\rho_p,\rho_e,\rho_\mu) &=&
 (\rho_n m_n +\rho_p m_p)
 + (\rho_n+\rho_p) \frac{E}{A}(\rho_n,\rho_p)
\nonumber\\ &&
 +\, \epsilon_\mu (\rho_\mu) + \epsilon_{el} (\rho_{el}),
\label{e:epsnn}
\eea
where we have used relativistic and ultrarelativistic approximations
for the energy densities of muons and electrons, respectively \cite{1986Shapiro}.
In practice, it is sufficient to compute only the binding energy of
symmetric nuclear matter and pure neutron matter,
since within the BHF approach it has been verified \cite{Baldo:1997ag,bombaci:1991zz} 
that a parabolic approximation for the binding
energy of nuclear matter with arbitrary proton fraction 
$x=\rho_p/\rho$, $\rho=\rho_n+\rho_p$,
is well fulfilled,
\be
 {E\over A}(\rho,x) \approx 
 {E\over A}(\rho,x=0.5) + (1-2x)^2 E_{\rm sym}(\rho) \:,
\label{e:parab}
\ee
where
the symmetry energy $E_{\rm sym}$ can be expressed in
terms of the difference of the energy per particle between pure neutron 
($x=0$) and symmetric ($x=0.5$) matter:
\be
  E_{\rm sym}(\rho) = 
   {{1\over 8} {\partial^2(E/A) \over \partial x^2}} {\Big |}_{x=0.5}
  \approx {E\over A}(\rho,0) - {E\over A}(\rho,0.5) \:.
\label{e:sym}
\ee
Once the energy density is known (Eq.~(\ref{e:epsnn})), the various chemical potentials 
(of the species $i=n,p,e,\mu$) can be computed straightforwardly,
\be
 \mu_i = {\partial \epsilon \over \partial \rho_i} \:,
\ee
and the equations for beta-equilibrium (\ref{e:beta}) allow one to determine the equilibrium composition $\{\rho_i\}$
at given baryon density $\rho$ and finally the EoS,
\be
 P = \rho^2 {d\over d\rho}
 {\epsilon(\{\rho_i(\rho)\})\over \rho}
 = \rho {d\epsilon \over d\rho} - \epsilon
 = \rho \mu_B - \epsilon \:.
\ee
\noindent

As far as the quark phase is concerning,  it is necessary to define the 
relations among the various $\mu_q$ that 
appear in Eq. (\ref{pquark}) and  the variable $\mu_B$.  
For this purpose we must distinguish two cases,
one with $\Delta=0$ and the other with $\Delta\neq 0$.
In the first case one has $P_{cfl}=0$ and the corresponding 
EoS for  quark matter is determined by the conditions
of  $\beta$-equilibrium and charge neutrality and baryon number conservation, 
as expressed by Eq.(\ref{e:beta}).  It is then sufficient to express each $\mu_q$ in terms of one 
single variable, namely $\mu_B$.
In the second case, with  $\Delta\neq 0$, the charge neutrality condition is realized in a 
peculiar way \cite{Rajagopal:2000ff}.  In fact, CFL pairing occurs if the  
number densities of the three flavors are equal
\be\label{equaldens}
\rho_u=\rho_d=\rho_s
\ee
which implies vanishing electron density, $\rho_e=0$, in order to 
maintain full charge neutrality. As explained in \cite{Rajagopal:2000ff},
Eqs. (\ref{equaldens}) with non vanishing  strange quark mass,
 $m_s\neq 0$,  are acceptable only if $\mu_e\neq 0$,
but in any case they  allow us to determine each single  
$\mu_q$ in terms of $\mu_B$.
Therefore, each time we consider the case with $\Delta\neq 0$,
we use the specific condition in Eqs. (\ref{equaldens}) to express 
$\mu_q$  in Eq. (\ref{pquark}) in terms of  $\mu_B$.  

Let us now discuss the main features of the hadron-quark phase transition,
which we assume to be first-order, thus performing  the Maxwell construction.
Fig.~\ref{fig:1} shows numerical results for the pressure as a function of the baryon chemical potential
$\mu_B$ in the hadronic matter and quark matter in beta equilibrium. In particular,
the green (red) solid curves represent the BHF (DBHF) EoS, whereas the remaining curves
are the results for the FCM model with different choices of the quark-antiquark potential $V_1$ (expressed in MeV).
For completeness, a negative value of the potential, $V_1=-50$ MeV is also included in this analysis. 
In the left, middle and right panels the value assumed for the gap $\Delta$ is respectively equal
to 0, 50 and 100 MeV.  All calculations shown in Fig. ~\ref{fig:1} are performed taking
$ G_2 = 0.006$  GeV$^4$. We notice that 
with increasing the value of $V_1$  the transition point is shifted to larger values of the chemical
potential, hence of the baryon density. However, the exact value depends also on the stiffness of the 
hadronic EoS at those densities. In this case, being the DBHF EoS stiffer than the BHF, the transition
takes place at smaller values of the density. We notice that 
the transition point is affected also by the value of the gap
$\Delta$, and is shifted toward smaller $\mu_B$ for larger value of the gap. 
We also see that no phase transition occurs
for negative values of $V_1$.

The resulting EoS, for the several cases discussed, is displayed in Fig.~\ref{fig:2}, where 
one can directly read off the phase transition between hadron matter and quark
matter under the Maxwell construction. We notice that the phase transition is allowed only for
$V_1 \geq 0$, and that the width of the plateau is directly related to $\Delta$. 
In the case $V_1 = 0$ (magenta curve) the phase transition takes place at very low value
of the density, the plateau is quite small and the pure quark matter phase starts at 
density about 3 times normal nuclear matter density. By increasing $V_1$ the 
phase transition is shifted to larger values of the energy density. 

\begin{figure*}
\centering
\resizebox{0.85\textwidth}{!}
{\includegraphics{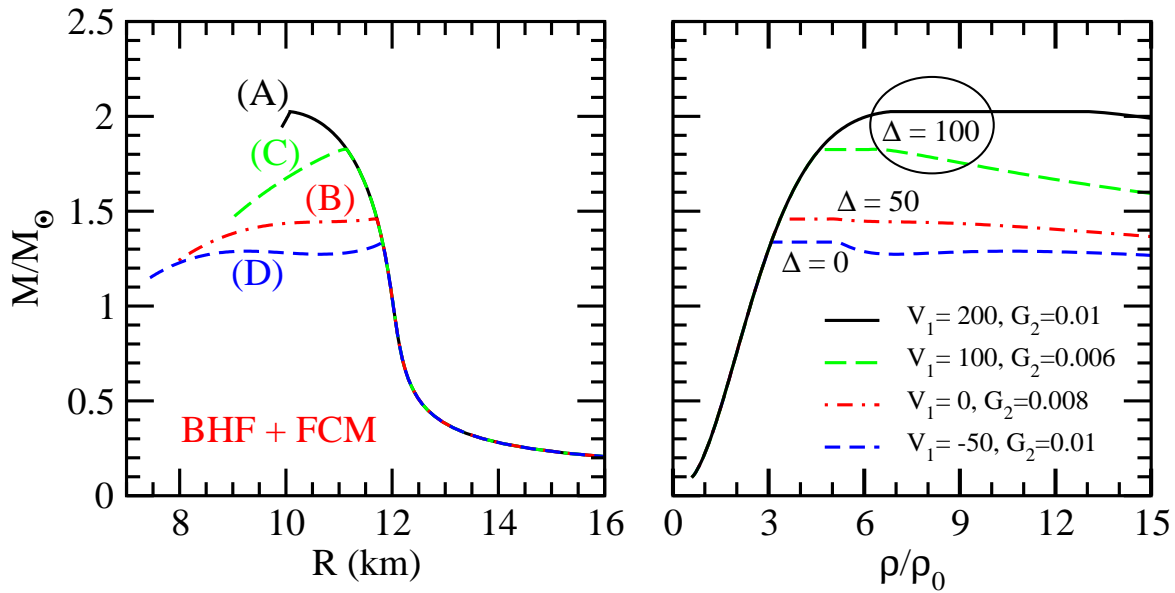}}
\caption{(Color online) The mass as function of the radius (left panel) and the central density (right panel) is displayed
for several values of $V_1$, $G_2$ and $\Delta$. The BHF EoS is used for the hadronic phase. The labels (A), (B), (C)
and (D) indicate the specific topologies of the hybrid star branch.}
\label{fig:3}       
\end{figure*}

\begin{figure*}
\vspace{15mm}
\centering
\resizebox{0.85\textwidth}{!}
{\includegraphics{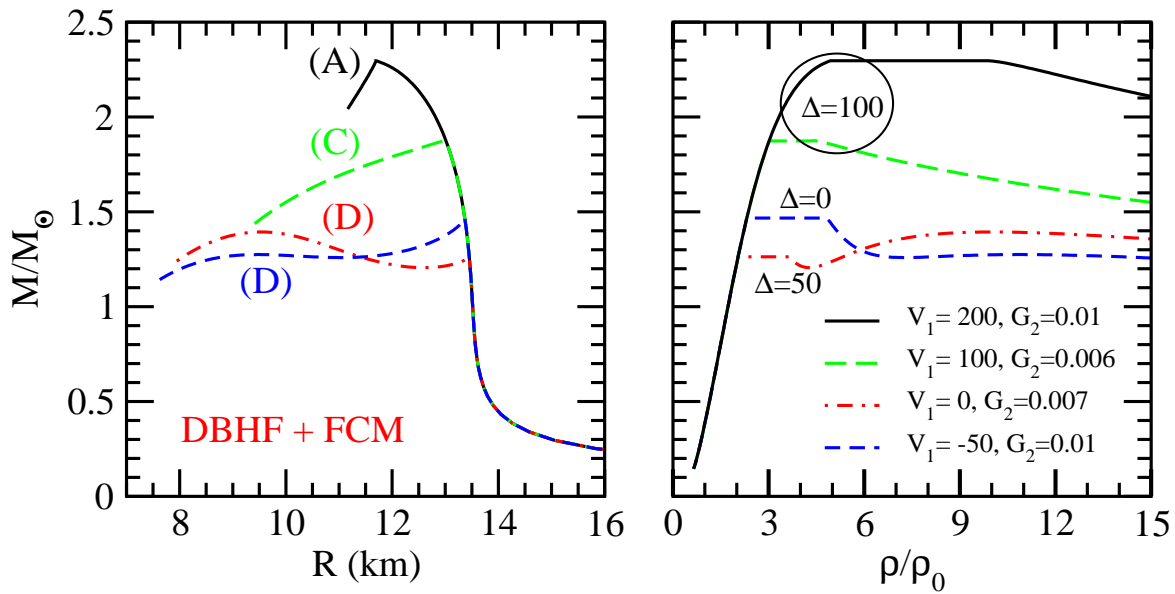}}
\caption{Same as Fig.3, but for the DBHF EoS.}
\label{fig:4}       
\end{figure*}

\subsection{Hybrid star structure}
\label{sec:4_2}

Once the EoS of the hybrid star matter is known,
one can use the Tolman-Oppenheimer-Volkoff \cite{1986Shapiro} equations
for spherically symmetric NS:
\bea
 \frac{dp}{dr} &=& -\frac{Gm\epsilon}{r^2}
 \frac{ (1+p/\epsilon) (1+4\pi r^3p/m) }{1-2Gm/r} \:,
\\
 \frac{dm}{dr} &=& 4 \pi r^{2}\epsilon \:,
\eea
where $G$ is the gravitational constant and
$m(r)$ is the enclosed mass within a radius $r$.
Given a starting energy density $\epsilon_c$,
one integrates these equations until the surface $r=R$,
and the gravitational mass is obtained by $M_G=m(R)$.
The EoS needed to solve the TOV equations is taken from the
neutron star matter calculations discussed above,  and matched with the crust EoS,
which has been taken from Refs.~\cite{1973NV}.

As is well known, the mass of the NS has a maximum value as a function of radius
(or central density), above which the star is unstable against collapse to a black hole.
The value of the maximum mass depends on the EoS,
so that the observation of a mass higher than the
maximum mass allowed by a given EoS simply rules out that EoS. 
This is illustrated in Fig.~\ref{fig:3},
where the relation between mass and radius (left panel)
and central density (right panel) in units of the saturation density $\rho_0=0.17~\rm fm^{-3}$
is displayed. Results are plotted for the case in which the BHF EoS is used for hadronic matter.
In Fig.~\ref{fig:3} we plot several cases obtained for different values of $V_1$, $G_2$ and $\Delta$ and
in this example the largest value of the maximum mass is observed for large values
of $V_1=200$ MeV, $\Delta=100$ and $G_2 = 0.01$, 
and it  is compatible with the largest mass observed up to now, i.e.  $(2.01\pm0.04) M_\odot$ 
in PSR J0348+0432 \cite{Antoniadis:2013en}.

Recently, it has been shown in ref. \cite{2015FCM} that the FCM equation of state can be accurately 
represented by the so-called CSS parameterization. The basic ansatz is that
a sharp phase transition occurs to a high-density phase, where the speed of sound is density-independent. 
As already discussed in ref.\cite{Alford:2013aca}, in all models of nuclear/quark matter one can find the 
four topologies of the mass-radius curve for compact stars: the hybrid branch
may be connected to the nuclear branch (C), or disconnected
(D), or both may be present (B) or neither (A). We will discuss in detail the FCM mapping onto the 
CSS parameterization in the next subsection. Here we limit ourselves to use the same
labels  in Figs.~\ref{fig:3} and ~\ref{fig:4}  in order to indicate the topology of the mass-radius curve,
which is strongly related to the values of the pressure and energy density at the transition point,
and to the energy density discontinuity. In the framework of the FCM model, 
the topology is related to the chosen values of $V_1$, $G_2$ and $\Delta$.
Using Fig.~\ref{fig:1} as a guide, we can obtain various  topologies just changing
$V_1$, $G_2$ and $\Delta$. For example,  when combining FCM quark matter to the BHF nuclear 
matter we find that,  for unpaired quark matter and $V_1 = \rm -50~MeV$, the lowest transition point can be obtained 
when  $G_2 > 0.006$  GeV$^4$. In Fig.~\ref{fig:3} the corresponding mass-radius relation, obtained with $G_2 = 0.01$  GeV$^4$
is displayed by the blue dashed line, which exhibits a branch of stable hybrid stars 
disconnected (D) by the hadronic branch.
With increasing $V_1$  the transition point moves
to larger values of the pressure and the energy density, and as a consequence we explore 
regions of the phase diagram where the topology changes. 
For instance, for $V_1=0$ we can get both (B) connected and disconnected hybrid star branches, 
whereas for $V_1=100~\rm MeV$ connected (C) hybrid star branches are present and, for the largest value
of $V_1=200~\rm MeV$ the hybrid branch is absent (A). This is clearly shown by a cusp in the
mass-radius relation,  and all configurations with radii smaller than the one characterizing the cusp are unstable.
Therefore only purely nucleonic stars do exist in this case. However, the stability
of those hybrid star configurations is related to the modeling of the deconfinement
phase transition, as pointed out in ref.\cite{2013Bom_Log} where the Gibbs construction was used instead of the Maxwell method.
The additional contribution of the CFL pressure to the FCM EoS produces only a shift of the transition point,
and therefore the topology explored can be different than the one of the unpaired case,
leaving unchanged the phase diagram. 

\begin{figure*}
\centering
\resizebox{0.85\textwidth}{!}
{\includegraphics{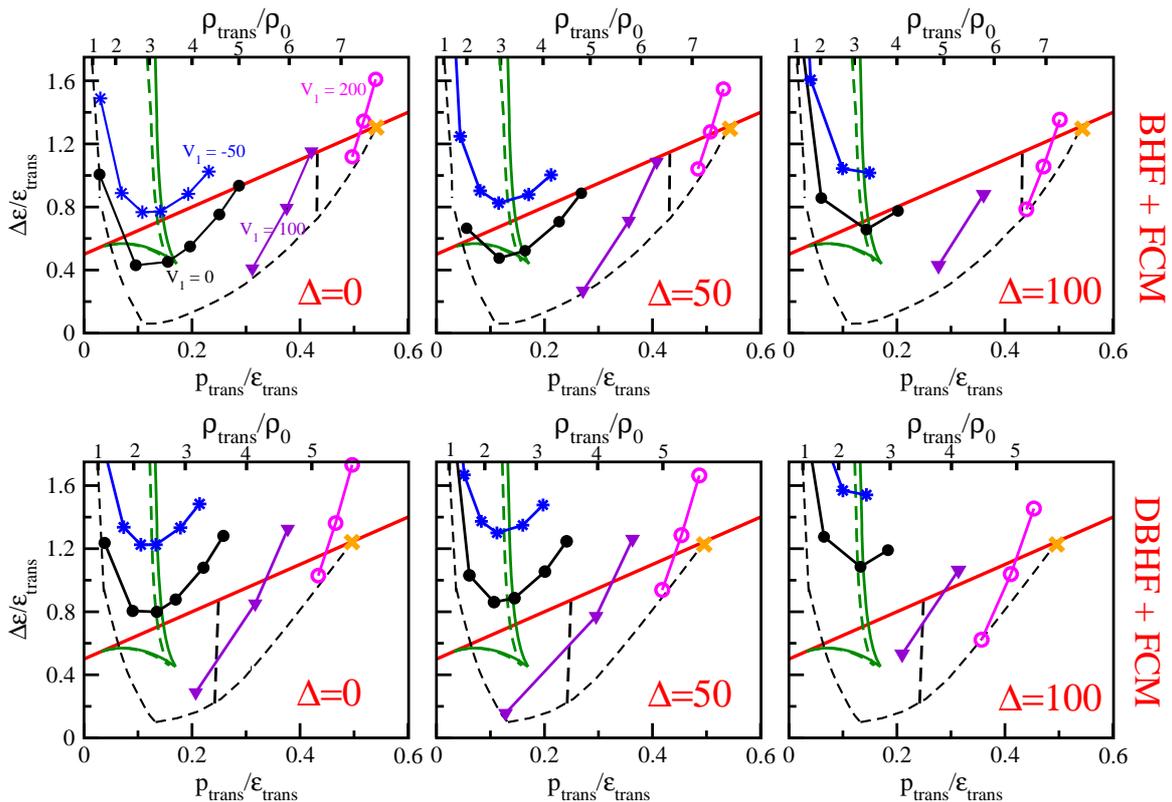}}
\caption{The mapping of the FCM quark matter model onto the CSS
parameterization.  Results are obtained using the BHF (upper panels) and DBHF
(lower panels) nuclear matter EoS.  The green curves are the
phase boundaries for the occurrence of
connected and disconnected hybrid branches. The dashed black line delimit the region yielded by the
FCM model. Within that region, the symbols give CSS parameter values for
FCM quark matter as $G_2$ is varied at constant $V_1$ (given in MeV).
The (orange) cross denotes the EoS with the  highest 
$p_{trans}$, which gives the heaviest FCM hybrid star.
The left, middle, and right panels display results obtained with $\Delta = \rm 0, ~50, ~and~ 100$ MeV respectively.}
\label{fig:5}      
\end{figure*}

In Fig.~\ref{fig:4} we display the mass-radius (left panel) and the mass-central density relation (right panel)
in the case that the EoS used for the hadronic phase is the DBHF.  We observe a topology 
similar to the one displayed in Fig.~\ref{fig:3}, except the (B) configurations, which do not appear for the chosen 
set of values used for $V_1$, $G_2$ and $\Delta$.  

Finally we comment on the values of the maximum mass. In both cases, either BHF or DBHF EoS for the
hadronic matter, we see that the largest possible values of the maximum mass are obtained only
for values of $V_1 > 100~\rm MeV$, and that only in the DBHF case maximum masses 
well above the observational limit are possible. In fact, the heaviest BHF+FCM hybrid star has a mass of 
$2.03~M_\odot$, and the heaviest DBHF+FCM hybrid star has a mass of $2.31~M_\odot$.
Those values are indicated by an orange cross in Fig.~\ref{fig:5}, where
we display the mapping between the FCM and CSS parameters which are,
in addition to the particular constant  value of the speed of sound $c_{QM}$,
the ratio of the pressure and energy density in nuclear matter
at  the transition point, $p_{trans}/\epsilon_{trans}$, and the ratio 
of the energy density discontinuity and the energy density at the transition, $\Delta\epsilon /\epsilon_{trans}$.
In the upper (lower) panels we show results for the BHF (DBHF) hadronic EoS, whereas in the left, middle and 
right panels calculations are reported for different values of the gap $\Delta = 0, ~50, ~100$ MeV respectively. 
The solid red line shows the threshold value $\Delta\epsilon_{crit}$ below which there is always a stable
hybrid star branch connected to the neutron star branch. This critical value is given by \cite{Haensel:1983,Seidov:1971,Lindblom:1998dp}

\begin{equation}
\frac{\Delta\epsilon_{crit}}{\epsilon_{trans}} = \frac{1}{2} + \frac{3}{2} \frac{p_{trans}}{\epsilon_{trans}}
\end{equation}
\noindent
and is obtained by performing an expansion in powers of the size of the core of the high-density phase. 
That result is analytical and  independent on both $c^2_{\rm QM}$ and the hadronic EoS. 
The solid (dashed) green lines represent the phase boundaries for connected and disconnected branches, and
are obtained for $c^2_{\rm QM}=1/3 (0.28)$. Those values span the
range of $c^2_{\rm QM}$ relevant for the FCM, as discussed in \cite{2015FCM}.  
\begin{table}[tbp]
\caption{
The minimum and maximum values of $G_2$ (in units of $\rm GeV^4)$ are shown for different
choices of $\Delta$ and $V_1$.}
\centering
\begin{tabular}{| c | c | c | c |}
\hline
 $\Delta$ (\rm MeV)   &    $V_1$ (\rm MeV)    &   $G_2^{min}$   &  $G_2^{max} $      \\
\hline
0.      & -50. & 0.007 & 0.014  \\
    & 0. & 0.005 & 0.012  \\
     & 100. & 0.003 & 0.01  \\
    &  200.  & 0.003 & 0.01  \\
\hline
100.      & -50. & 0.009 & 0.014  \\
    & 0. & 0.007 & 0.012  \\
     & 100. & 0.006 & 0.01  \\
    &  200.  & 0.003 & 0.01  \\
\hline
\end{tabular}
\label{t:G2}
\end{table}

\begin{table}[tbp]
\caption{
The total radius R, the radius of the quark core $\rm R_{Q}$, the radius of the hadronic layer $R_{H}$ and the crust radius $R_{crust}$ are given for a hybrid star mass $M=2 M_\odot$, for different choices of the hadronic EoS and $\Delta$. All radii are given in km.}
\centering
\begin{tabular}{| c | c | c | c | c | c |}
\hline
 EoS & $\Delta$ (\rm MeV)   &  R   &  $\rm R_{Q} $ & $\rm R_{H}$ & $\rm R_{crust}$     \\
\hline
BHF & 0.  & 10.37 & 0.055 & 9.97 & 0.345  \\
        & 100. & 10.44 & 0.215 & 9.87 & 0.355  \\
\hline
DBHF & 0. & 12.78   & 1.27 & 10.87 & 0.640  \\
          &  100.  & 12.72 & 2.42 & 9.665 & 0.635  \\
\hline
\end{tabular}
\label{t:Rad}
\end{table}
The dashed black contour delimit the region accessible by the FCM calculation. Above that region, 
the symbols connected by solid lines show the CSS parameterization of the FCM quark matter EoS.
Along each line we keep $V_1$ constant and vary $G_2$ over the range indicated in Table~\ref{t:G2} for the
two extreme cases $\Delta=0, ~100$ MeV and different values of $V_1$, for both BHF and DBHF.
In Fig.~\ref{fig:5} $V_1$ varies from -50 MeV up to the maximum value at which hybrid star configurations occur, 
which is indicated by an (orange) cross. For the BHF case that value is
$V_1=240\,\, \rm MeV$, $G_2=0.0024\,\,\ \rm GeV^{4}$ and for the  DBHF case 
it is $V_1=255\,\,\ \rm MeV$, $G_2=0.0019\,\,\ \rm GeV^{4}$.
The vertical black dashed lines indicate the parameter regions accessible by the FCM and consistent 
with the measurement of a $M = 2~M_\odot$. Hybrid stars with mass heavier than $2~M_\odot$ lie on a very small
connected branch on the right side of the vertical black dashed lines, and cover a small range of central 
pressures, having a very tiny quark core, with mass and radius similar to those of the heaviest 
purely hadronic star, as was already discussed in Ref.~\cite{2015FCM}.
For completeness, we display in Table~\ref{t:Rad} the characteristic radius of a hybrid star with 
$\rm M=2 M_\odot$ obtained with BHF and DBHF EoS for the hadronic phase and two extreme values
for $\Delta=0, 100$ MeV. We chose typical configurations lying on the vertical black lines plotted in Fig.~\ref{fig:5}.
We notice that the radius of the quark core $\rm R_{Q}$ is bigger for the stiffest hadronic EoS, being comprised
between 1 and 3 km, whereas for the soft hadronic EoS the quark core radius $\rm R_Q$  is not larger that a few hundreds meters. In both cases the hadronic layer occupies the largest portion of the star, and is characterized by a radius $\rm R_H$ of about 10 km. The crust radius $\rm R_{crust}$ is
always smaller than 1 km.

Moreover we notice in Fig.~\ref{fig:5}  that along each line of constant $V_1$, $p_{trans}/\epsilon_{trans}$ grows with $G_2$,
and this can be explained by recalling the linear dependence of the quark pressure on $G_2$ in 
Eq.~(\ref{pqgp2}), so that, at fixed chemical potential, 
an increase of $G_2$ lowers the quark pressure, making quark matter less favourable,
and shifting the transition point to higher chemical potential or pressure.
This was already discussed in Ref.~\cite{Baldo:2008en} for 
BHF nuclear matter, and is equally applicable to DBHF nuclear matter.
We also see in Fig.~\ref{fig:5}  that the combination of $G_2$ and $V_1$ moves the curves inside the region accessible by FCM which
is delimited downward by the dashed black line. 
Fig.~\ref{fig:5} shows that the introduction of a color-flavor locking effect characterized by a 
gap $\Delta$ does not change 
qualitatively the gross features of the phase transition, being the topology of the hybrid star
branch  slightly affected.

\begin{figure*}
\vspace{10mm}
\centering{
\resizebox{0.75\textwidth}{!}{%
\includegraphics{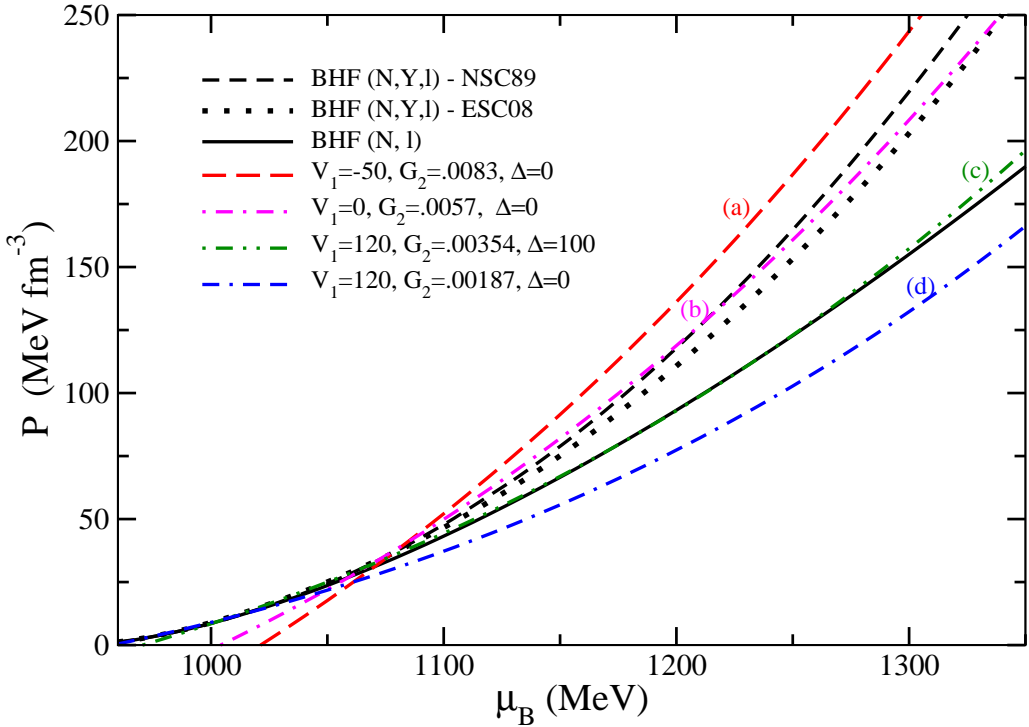}}
}
\caption{ Pressure vs. baryon chemical potential
 corresponding to the hadronic EoS's, including hyperons 
(dashed and dotted black lines) or without hyperons (BHF)  (solid black 
line),  and their crossing with the quark matter pressure evaluated 
with four different parameterisations ((a),(b),(c),(d)) lines).}
\label{fig:6}       
\end{figure*}

\subsection{ Effects of hyperons on the phase transition}
\label{sec:4_3}

It is known that the effect of including hyperons in the hadronic EoS is to soften the interior
of the NS so that it becomes difficult to get masses of the stars as heavy as 2 solar masses.
In our analysis, this is confirmed by the pressure obtained for the two parameterizations NSC89
and ESC08, which are plotted in Fig.~\ref{fig:6} (respectively, dashed and dotted black curves)
together with the BHF EoS (solid black curve) used in  Sect.~\ref{sec:4_1}.
In fact, the steeper growth with $\mu_B$ of the former two curves with respect to the latter is 
an indication of the greater stiffness of the EoS when hyperons are neglected.
At the same time we notice that the NSC89 and ESC08 parameterizations quantitatively give 
very close results. We stress that in this paper we discuss hypèeron effects obtained only in the BHF
approach.

Then, it is easy to realise that the inclusion of hyperons 
puts more stringent constraints on the parameters of the quark matter EoS, 
in order to observe a crossing of the pressures in the two phases.  In 
Fig.~\ref{fig:6} we report four examples of quark matter pressure for different choices
of the parameters (curves (a),(b),(c),(d)).  These curves explicitly show that the parameter
$V_1$ is mainly responsible for their slope, while $\Delta$ has a much smaller effect and 
$G_2$, being an additive constant to the pressure  as shown in  Eq. (\ref{pqgp2}), produces 
a global upward or downward shift of the quark matter pressure. 

Therefore, when going from 
$V_1=-50$ MeV to $V_1=0$ to $V_1=120$ MeV, the corresponding curves (a), (b) and (c)
become less and less steep,  and one can observe three representative behaviors:
(a) shows a phase transition at a crossing point below 1100 MeV with any hadronic EoS;
(b) has  the same crossing as in the case (a) and, when compared to the BHF EoS, it
remains the favoured phase at any $\mu_B$, but  when compared to the NSC89 or ESC08
parameterizations, one observes a second crossing at larger $\mu_B$; 
(c), after an interval in which the pressures for the hadronic and quark phases are substantially
the same, the NSC89 or ESC08 curves stay above  (c) which, in turn is above
the BHF curve. 

\begin{figure*}
\vspace{25mm}
\centering{
\resizebox{0.75\textwidth}{!}{%
\includegraphics{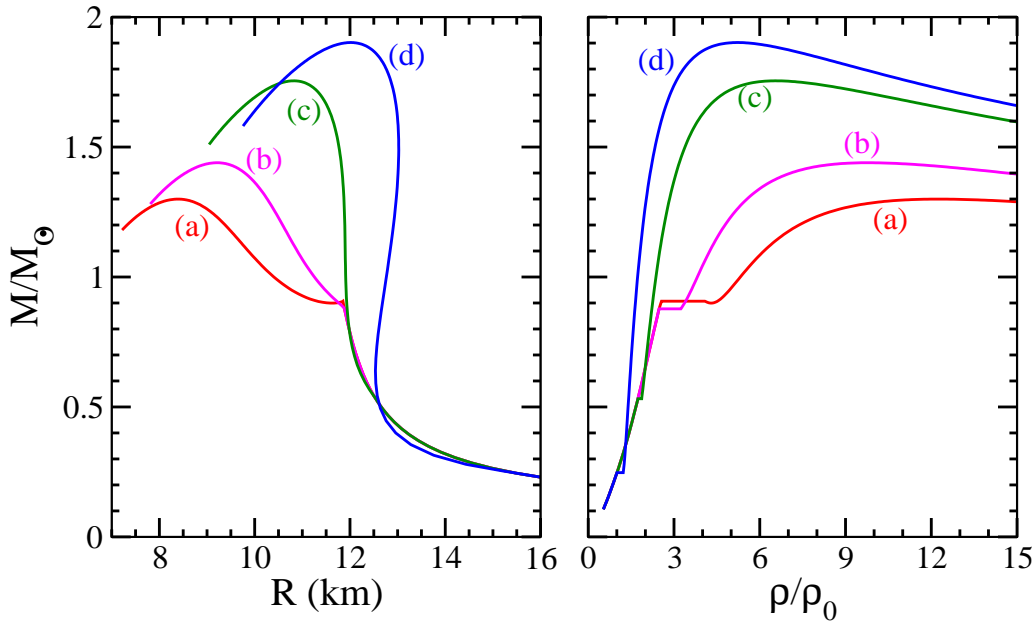}}
}
\caption{Mass-radius (left panel) and mass-central density (right panel) plots of the 
hybrid stars corresponding to the quark matter EoS (a), (b), (c), (d) 
together with the hyperon parameterization ESC08 of Fig.~\ref{fig:6}.}
\label{fig:7}       
\end{figure*}

Therefore, one learns that the quark pressure can exceed the  pressure of hadronic EoS including
hyperons only at small $V_1$, typically well below 100 MeV, and one knows from the analysis of 
Sect.~\ref{sec:4_2}  that smaller $V_1$ correspond to NS with smaller masses. 
To verify this point, we consider the cases (a) and (b)  (retaining for (b) only the first crossing 
of the quark and hadronic pressure)  and then  derive the corresponding mass-radius or 
mass-central density relation which are given  in  Fig.~\ref{fig:7} for the ESC08 EoS 
(the NSC89 case produces almost indistinguishable results).  
It is evident that in the cases (a) and (b), the masses remain below 
1.5 $\rm M_\odot$.  In addition, for (b) we ignored  the second crossing point above which,
in principle, the hadronic phase becomes again favourable, but, in any case, this new transition 
would make the NS even softer thus lowering its maximum mass and ruling out the possibility of 
reaching 2 $\rm M_\odot$.

Before concluding, we reconsider in the other cases 
the procedure followed for (b), namely the derivation of  the 
mass-radius relation obtained by  systematically  considering 
only  the  transition from hadronic to quark matter occurring at the 
lowest value of  $\mu_B$, while ignoring other potential  
transitions occurring at higher $\mu_B$. 

\begin{figure*}
\vspace{15mm}
\centering{
\resizebox{0.75\textwidth}{!}{%
\includegraphics{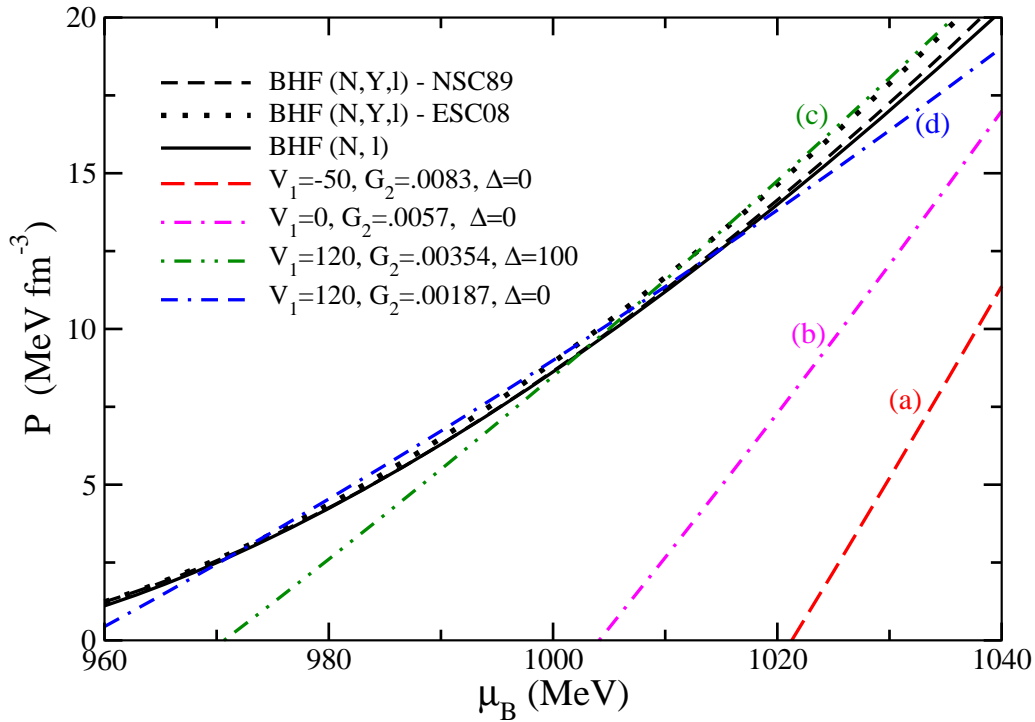}}
}
\caption{Enlargement of the low $\mu_B$ region of Fig.~\ref{fig:6}.}
\label{fig:8}     
\end{figure*}

By following this procedure, 
which is somehow justified by our poor knowledge of the hyperon 
interactions, we  reconsider in detail 
the various examples of Fig.~\ref{fig:6} and, for convenience, we 
report in Fig.~\ref{fig:8} an enlargement of the region at 
small $\mu_B$. In fact, from Fig.~\ref{fig:8} it is evident that the 
double  transition which is observed at a larger scale in Fig.~\ref{fig:6} 
for the case (b), now occurs on a smaller scale for the cases (c) and (d).
These two examples are realized with $V_1=120$ MeV and, respectively, 
$\Delta=100$ MeV and $\Delta=0$ and $G_2$ is tuned to make the quark matter
pressure almost tangent to the NSC89 or ESC08.  While the  case (c) crosses for the first time
the hadronic EoS's  slightly above $\mu_B\sim 1000$ MeV, where  the ESC08 and the BHF EoS's
are distinguishable, the crossing of curve (d) occurs below $\mu_B\sim 980$ MeV, before the onset
of hyperons. 

In both cases (c) and (d), the second crossing is very close, but according to the assumption made we retain only the 
quark matter EoS at larger $\mu_B$ after the first crossing.  The interesting point is that, as $V_1 =120$ MeV 
is rather large for these two cases, one  expects large maximum masses for the corresponding NS.
This can be verified by looking at the  right panel of Fig.~\ref{fig:7}, where after a very small plateau, related to the 
small difference between the quark and hadronic pressure along a rather large interval around the transition point,
the mass of the NS grows above 1.7 $M_\odot$ for (c) and up to 1.95 $M_\odot$ for (d)
that is reasonably  close to the observational constraint of 2 $M_\odot$. 
It is remarkable that very similar results are obtained for the 
maximum NS masses in \cite{Kurkela:2009gj} where the 
NSC89 parameterization is used for hyperons and a sort of  
QCD corrected bag model for quark matter.

\section{ Conclusions}
\label{sec:5}

The FCM extension at finite $T$ and $\mu_B$ provides us a very simple description 
of the quark dynamics in terms of two parameters, namely the gluon condensate $G_2$, that parametrizes 
the vacuum pressure and energy density, and hence is strictly related to the bag constant of the MIT bag model, 
and the potential $V_1$, which summarizes the interaction corrections to the free quark and gluon pressure.
It is then natural to explore the predictions of the FCM on the maximum masses (and their associated radii) 
of NS when these parameters are varied.
In order to have a more complete picture, we include the effect of color superconductivity through the CFL mechanism,
which amounts to the addition of a new free energy contribution written in terms of the gap $\Delta$.

Clearly, any prediction of a quark matter model on the structure of a hybrid NS strongly depends on the 
nuclear matter EoS employed  and among the 
large variety of nuclear EoS available in the literature, we focused on the non-relativistic BHF EoS and its relativistic counterpart,
the DBHF EoS. These are derived within a solid microscopic approach, and give us different predictions on 
the NS maximum mass. For completeness, we also analyzed the inclusion of the hyperon degrees of freedom 
that produces a softening of the nuclear matter EoS with the consequent reduction of the maximum mass of the NS.

With this new set of more refined calculations, we confirm the trend already observed in \cite{Baldo:2008en,Plumari:2013ira,2015FCM}, 
i.e. the maximum mass of hybrid stars grows 
with the two parameters $V_1$ and $G_2$ while it decreases when $\Delta$ is increased. 
More interestingly, we extend the mapping developed in \cite{2015FCM} among the parameters of the FCM and those defining the CSS parameterization, by displaying
the effect of the gap $\Delta$. In fact, from the various panels of Fig.~\ref{fig:5}, it is evident that the border of the area of the diagram 
accessible by the FCM (i.e. the dashed black curve) is not sensitive to $\Delta$ and it is totally determined in terms of the 
CSS parameters. In particular, even the region corresponding to configurations associated to hybrid stars with 
maximum mass greater than 2 $M_\odot$, which is the triangle-like area delimited above by the straight solid red line, 
below by the dashed black curve and finally on the left by the almost vertical dashed black segment, is only determined in terms 
of the two CSS parameters reported on the $x$ and $y$ axes of the diagrams.
Therefore one can conclude that a particular configuration with mass around or above two solar masses 
can be realized in the FCM by different pairs of $G_2$ and $V_1$, depending on the specific value
assigned to $\Delta$, i.e. 
the appearance of a color superconducting gap can be mimicked by a shift of the other two parameters.
Therefore, even the mass of the heaviest hybrid star predicted by the FCM (the orange crosses in Fig.~\ref{fig:5}) 
does not correspond to a unique set of $G_2$, $V_1$ and $\Delta$, while, as seen in \cite{2015FCM},
its value strongly depends on the specific choice made for the nuclear matter EoS.

The inclusion of the hyperons induces dramatic changes in this picture. In fact, a regular transition
from nuclear to quark matter with a stable quark phase up to very high chemical potential requires a particular tuning 
of the FCM parameters that leads to very low maximum masses, below 1.5 $M_\odot$.
We have also observed that it is possible to find specific sets of the parameters $G_2$, $V_1$ and $\Delta$
such that the hadronic and quark matter pressure run very close for a large range of $\mu_B$ and, when 
looking more in detail, one observes multiple crossings of these lines, although at large $\mu_B$ the 
phase that includes hyperons is favoured. 
For completeness we analyze these cases by retaining only the first crossing from the hadronic to the quark matter phase,
and neglecting the other transitions at higher chemical potential.
In this case it is possible to tune the FCM parameters in such a way to raise the maximum mass up to 1.95 $M_\odot$, although 
an explanation supporting the strong assumption on the transition is required in order to accept this result.

\section*{Acknowledgments}
We thank H.-J. Schulze (INFN Sezione di Catania) for providing us with
the BHF EoS for hypernuclear matter with the NSC89 and ESC08 nucleon-hyperon 
parameterizations. We are also grateful to M. Alford for bringing the CSS parameterization to our attention.
Partial support comes from ``NewCompStar'', COST Action MP1304. 

\newcommand{\aap}{Astron. Astrophys.\ }

\bibliographystyle{JHEP_MGA}
\bibliography{fcm} 

\end{document}